\documentclass[10pt,a4paper,fleqn]{article}

\usepackage[
    top=20mm,
    bottom=20mm,
    left=18mm,
    right=18mm
]{geometry}

\usepackage{graphicx}
\usepackage{multicol}
\usepackage{titlesec}
\usepackage{fancyhdr}
\usepackage{setspace}
\usepackage{array}
\usepackage{booktabs}
\usepackage{tikz}
\usepackage{tcolorbox}
\usepackage{amssymb}
\usepackage{caption}
\usepackage{fontspec}
\usepackage{amsmath}
\usepackage{graphicx}
\usepackage{siunitx}
\usepackage{multirow}

\usepackage[switch]{lineno}
\usepackage[numbers,sort&compress]{natbib}

\usepackage{hyperref}
\usepackage{url}
\usepackage{lipsum}

\usepackage{booktabs}
\usepackage{titlesec}
\titleformat{\section}
    {\normalsize\bfseries}
    {\thesection}
    {1em}
    {}

\titleformat{\subsection}
  {\footnotesize\bfseries}
  {\thesubsection}
  {1em}
  {}

\titleformat{\subsubsection}
  {\footnotesize\bfseries}
  {\thesubsubsection}
  {1em}
  {}

\titleformat{\paragraph}[runin]
  {\footnotesize\bfseries}
  {\theparagraph}
  {1em}
  {}

\usepackage{lipsum}

\usepackage{varwidth}
\usepackage{adjustbox}

\definecolor{tether-gray}{HTML}{F2F1EF}

\newtcolorbox[auto counter]{custombox}[2][]{
    title={Box~\thetcbcounter: #2},
    label={#1},
    fonttitle=\bfseries,
    coltitle=black,
    before title=\vspace{15pt},
    colback=tether-gray,
    colframe=tether-gray,
    boxrule=0pt,
    arc=5pt,
    left=15pt,
    right=15pt,
    bottom=15pt,
    width=\textwidth,    
}

\begin{document}




\begin{flushleft}
\includegraphics[height=20pt]{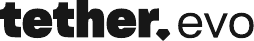}
\end{flushleft}

\vspace{10pt}


\begin{tcolorbox}[
    colback=tether-gray,
    colframe=tether-gray,
    boxrule=0pt,
    arc=5pt,
    left=15pt,
    right=15pt,
    top=15pt,
    bottom=15pt,
    width=\textwidth
]


{\huge\bfseries
Retrieval-Based Brain Decoding by \\ Alignment, not Complexity
}

\vspace{5pt}


{\small\bfseries
Matteo Ciferri\textsuperscript{1},
Matteo Ferrante\textsuperscript{1,2},
Nicola Toschi\textsuperscript{1,3}
}

\vspace{2.5pt}

{\small
\textsuperscript{1}Department of Biomedicine and Prevention, University of Rome Tor Vergata, Rome, Italy\\
\textsuperscript{2}Tether Evo\\
\textsuperscript{3}A.A.
Martinos Center for Biomedical Imaging, Massachusetts General Hospital, Boston, USA\\
}

\vspace{15pt}


{\footnotesize
\textbf{Abstract}. A prominent theory in cognitive science suggests that concepts in the brain are organized as high-dimensional vectors, with semantic meaning captured by directions and relative angles in this space. Brain decoding is the effort of reconstructing or retrieving stimuli (or their representations) from neural activity and involves finding a function that approximates how the brain represents concepts. This motivates the investigation of contrastive objectives as biologically plausible candidates to reverse the brain loss function. In this work, we study how functional MRI (fMRI) activity can generally be mapped with the embedding spaces of foundation models in vision, language, and audio. Although neural computations are highly non-linear at the microscale, fMRI measurements average signals across space and time, further smoothed by noise, effectively linearizing the observable representation. Consistent with these views, our experiments across multiple datasets demonstrate that linear contrastive decoders consistently outperform ridge regression and standard non-linear alternatives, and that these results generalize across images, text, and sound. These findings indicate that decoding gains arise more from the choice of training objective than from architectural complexity, pointing to contrastive-linear models as a principled strategy for brain decoding.
}

\end{tcolorbox}

\vspace{10pt}


\footnotesize


\section{Introduction}


A central challenge in cognitive science is to understand how the brain represents concepts and encodes sensory information. Recent theoretical work argues that human concepts are most plausibly represented as high-dimensional vectors \citep{whyconcepts2024}. Vector-based representations naturally explain typicality and similarity effects through distances in the representational space, capture relations and analogies via vector arithmetic, support compositionality and theory-like structures, and even allow the flexible formation of ad hoc categories. This framework unifies different theory-based views of concepts under a single representational format. Moreover, it resonates with the success of modern foundation models, which learn rich embedding spaces where meaning is encoded in geometric relations among vectors. If the brain organizes concepts in such vector spaces, then comparison and learning are likely driven by similarity, suggesting that contrastive learning provides a biologically plausible approximation of the brain’s own optimization principle. 

At the same time, a complementary line of research has challenged the assumption that modeling brain dynamics at the macroscale necessarily requires complex non-linear systems. A recent large-scale study from \cite{Nozari2024} on fMRI data showed that linear models not only match but often outperform a wide range of non-linear approaches across predictive accuracy, residual structure, and computational efficiency. The authors traced this apparent linearity to several factors intrinsic to macroscopic measurements: spatial averaging over millions of neurons, temporal filtering of fast dynamics, observation noise, and the limited sample size relative to dimensionality. Together, these effects act to smooth and linearize the measured signal, such that what fMRI captures is effectively a first-order approximation of the underlying neural computations. This provides a principled explanation for why linear models can be highly effective in fMRI decoding, despite the non-linear nature of the neural processes they ultimately reflect.

Building on these perspectives, this work is motivated by several key questions: What is the most effective way to map brain activity into the embedding spaces of foundation models? Do more complex non-linear models provide an advantage in the context of noisy and high-dimensional data? Is decoding performance driven more by vector alignment in the representational space than by average error minimization?

To address these questions, we systematically study brain decoding from functional Magnetic Resonance Imaging (fMRI) across three distinct modalities (images, music, and text), using embeddings extracted from state-of-the-art foundation models. We evaluate a spectrum of decoding models, ranging from ridge regression to linear mappings trained with contrastive loss, to shallow MLPs. Our findings reveal three key insights: (i) a simple linear mapping trained with contrastive learning consistently outperforms ridge regression across modalities; (ii) introducing non-linearities via MLPs does not improve decoding performance, and in fact degrades retrieval accuracy; (iii) prioritizing discriminative separation of embeddings is more important than minimizing pointwise error.

We emphasize that our conclusion is related to the data regime and preprocessing commonly used in fMRI decoding (GLM betas or HRF-averaged responses). We do not claim that non-linear architectures cannot outperform linear ones in other settings—for example, with minimally averaged data, larger datasets, or temporal models.

\section{Related Works}

Recent years have witnessed remarkable progress in decoding complex stimuli from neural activity, particularly in non-invasive settings such as fMRI \citep{nishimoto2011reconstructing,huth2016natural,ferrante2023eyes,banville2025scalinglawsdecodingimages}. 
In the visual domain, approaches leveraging pre-trained vision–language models such as CLIP, combined with linear regression or contrastive learning, have enabled retrieval-based decoding and even realistic image reconstruction when coupled with diffusion models \citep{ozcelik2023natural,lin2022mind,scotti2023reconstructing,xia2023dream}. Beyond vision, growing evidence shows that fMRI activity can be mapped onto latent spaces of diverse modalities—including video \citep{chen2023cinematic}, language and music \citep{tang2023semantic,jalouzot2025optimizing,denk2023brain2musicreconstructingmusichuman,ciferri2026r,ye2025generative}. The advent of large pre-trained models has been a key enabler of this progress, providing rich representational spaces that support both retrieval tasks and generative reconstruction from neural data. 

Brain decoding approaches have traditionally relied on linear methods such as ridge regression to predict high-dimensional representations of stimuli from non-invasive neural recordings \citep{ozcelik2023natural,liu2023brainclip,denk2023brain2musicreconstructingmusichuman}. While successful in controlled settings, these models are limited in their ability to capture the semantic richness of natural stimuli. \textcolor{black}{More complex neural networks have shown promising results \citep{scotti2024mindeye2,xia2024umbrae,careil2025dynadiff}, and several works have employed contrastive learning within deep architectures for brain encoding/decoding \citep{ciferri2026mapping,scotti2023reconstructing,chen2022seeing}. However, it is unclear whether performance gains stem from the contrastive objective itself or from architectural complexity. Our work addresses this directly: we provide a systematic and controlled comparison across decoder types under matched conditions, across three input modalities and corresponding foundation models. To our knowledge, no prior work has conducted this kind of benchmarking, isolating the contribution of the training objective from that of architectural depth. Furthermore, most studies have focused on a single modality, leaving open the question of whether decoding strategies generalize across different types of cognitive data.}


\section{Material \& Methods}

\begin{figure*}
    \centering
    \includegraphics[width=1\linewidth]{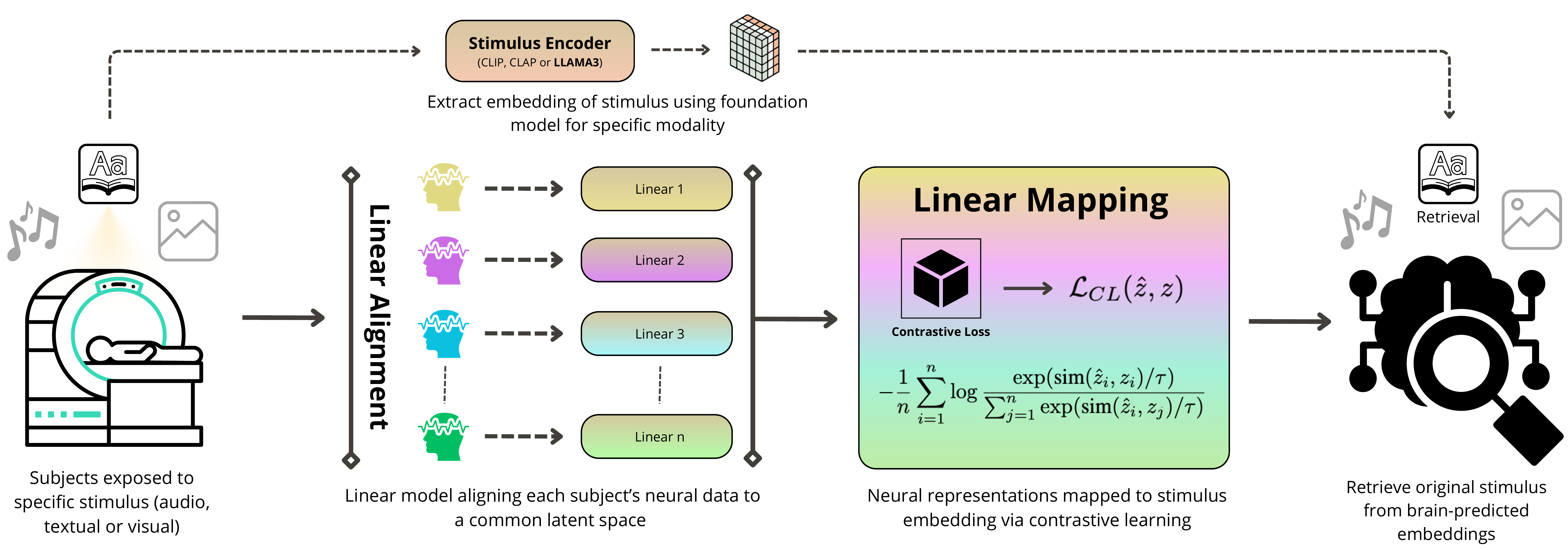}
    \caption{The same linear contrastive model is employed across three experimental conditions, differing only in the stimulus modality (audio, textual, or visual). 
    For each modality, neural responses from fMRI are aligned through subject-specific linear transformations and mapped into the corresponding stimulus embedding space (obtained from a pretrained foundation model such as CLIP for images, CLAP for audios, or LLaMA for text) via a contrastive learning objective. Training is carried out independently for each modality. At test time, retrieval is performed by comparing brain-predicted embeddings with estimated stimulus embeddings.}
    \label{fig:pipeline}
\end{figure*}

An overview of the proposed framework is shown in Figure~\ref{fig:pipeline}. \textcolor{black}{Our approach learns a mapping from neural responses to the embedding space of a pretrained foundation model, enabling retrieval of the corresponding stimulus from brain activity alone. Each modality is trained and evaluated independently, with its own dataset, ROI definition, and target embedding space; we do not train a joint cross-modal decoder. The same model architecture is employed across all modalities, demonstrating that our framework achieves consistent improvements over ridge regression and non-linear alternatives.} The following subsections provide details on the datasets, the decoder design, the training objective, and the evaluation metrics.

\subsection{Image Processing}

For the visual dataset, we relied on the \textit{Natural Scenes Dataset} (NSD) from \cite{Allen2022}, which includes fMRI data acquired while multiple subjects viewed natural images. The dataset provides over 24,000 fMRI trials: a training set of approximately 8,859 unique images per subject, and a shared set of 982 images viewed by all subjects, held out exclusively for evaluation. Decoder fitting and hyperparameter selection used training data only. We focused on data from Subj01, Subj02, Subj05, and Subj07, normalized and transformed into MNI space at 2~mm resolution. fMRI signals were preprocessed using anatomically predefined NSD General ROI masks and beta coefficients estimated via GLM \citep{glmdenoise, glmsingle}. Each neural sample is represented as a vector of 15,724 voxels corresponding to the estimated beta responses within these ROIs. For each stimulus, the corresponding image is fed into CLIP \citep{clip} at inference mode, yielding a 512-dimensional embedding from the image-text projection layer.

\subsection{Language Processing}

We used the publicly available dataset from \cite{lebel2023natural}, focusing on three subjects (S1, S2, S3), hereafter referred to as HUTH Language. Each subject underwent approximately 16 hours of fMRI recordings while listening to 83 naturalistic stories from \textit{The Moth} and \textit{Modern Love} podcasts. fMRI data were acquired with a 3T scanner (TR = 2.00~s, voxel size = 2.6~mm isotropic). Standard preprocessing included motion correction, cross-run alignment, standardization, and removal of low-frequency drifts. The first 70 stories were used for training, 12 for validation, and the story \textit{wheretheressmoke} (250 sentences, repeated 10 times) served as the test set.
Language-sensitive voxels were identified via an encoding model mapping LLaMA3-8B (layer 13) word embeddings \citep{dubey2024llama3herdmodels} to fMRI responses. Word embeddings were computed using a five-word context window, matching \cite{tang2023semantic} for comparability across decoders, and downsampled with a Lanczos filter to match the fMRI temporal resolution. The top 10,000 cortical voxels with highest encoding model correlation on held-out data were selected as decoding inputs \citep{tang2023semantic, toneva2019interpreting}.

\subsection{Music Processing}

We employed the GTZan fMRI dataset \cite{NAKAI2022107675}, consisting of recordings from five participants (sub-001 to sub-005) who each listened to 540 music excerpts across ten genres. Each stimulus lasted 15 seconds (22.05~kHz, 2-second fade-in/out), with 18 runs per subject (12 training, 6 testing) of 40 clips each. fMRI data were acquired at TR = 1.5~s (3.0T scanner). Preprocessing included motion correction, MNI co-registration via FSL, detrending, run-wise standardization, and hemodynamic delay correction (first 3 TRs discarded). Neural responses were averaged over the following 10 TRs per stimulus, yielding 480 training and 60 test pairs per subject.
Music-responsive voxels were identified via voxel-wise regression models predicting brain activity from CLAP \citep{elizalde2022clap} embeddings (512-dimensional, audio-text projection layer), with regularization optimized through nested cross-validation. We retained the top 3,000 voxels ranked by Pearson correlation on held-out training data, yielding subject-specific masks of music-responsive regions used as input for decoding.

\subsection{Neural-to-Embedding Decoder}
\label{subsec:mapping_model}

We designed a neural architecture to learn a mapping between neural activity and the target embedding space using a contrastive learning framework. The best architecture is simple: the decoder is composed of a sequence of linear layers, so that the transformation is essentially a stack of affine projections. 

Formally, given an input vector $\mathbf{x} \in \mathbb{R}^{d}$ and a subject index $k$, the model first applies a subject-specific alignment layer $A_k$, followed by hidden projections and an output layer:
$\mathbf{z} = W_o \, W_h \, A_k \mathbf{x},$
where $A_k \in \mathbb{R}^{d \times d_c}$ aligns the subject-specific input to a shared dimensionality $d_c$, 
$W_h \in \mathbb{R}^{d_c \times h}$ projects to hidden dimension $h$, 
and $W_o \in \mathbb{R}^{h \times d_o}$ maps to the output embedding space of foundation model $d_o$. 

In order to align neural representations with target embeddings $\mathbf{y} \in \mathbb{R}^{d_o}$, 
we employ a contrastive loss inspired by the NT-Xent formulation. 
For a batch of predicted embeddings $\{\mathbf{z}_i\}_{i=1}^{N}$ and targets $\{\mathbf{y}_i\}_{i=1}^{N}$, 
the cosine similarity is computed as
$S_{ij} = \frac{\mathbf{z}_i^\top \mathbf{y}_j}{\|\mathbf{z}_i\|\|\mathbf{y}_j\|}.$
The loss encourages each $\mathbf{z}_i$ to be most similar to its paired $\mathbf{y}_i$:
\[
\mathcal{L}_{\text{contrastive}} = - \frac{1}{N} \sum_{i=1}^{N} 
\log \frac{\exp(S_{ii}/\tau)}{\sum_{j=1}^{N} \exp(S_{ij}/\tau)},
\]
where $\tau > 0$ is a temperature hyperparameter. 

The decoder is trained end-to-end with AdamW optimization, using early stopping based on validation loss. \textcolor{black}{As baselines, we compared against (i) multivariate ridge regression from the full voxel vector to the target embedding vector, trained with L2 loss, and (ii) an MLP decoder with non-linear activations. All three decoders output vectors in the same target embedding space and are evaluated with the same cosine-similarity retrieval protocol.}

\subsection{Evaluation}

At test time, we run the decoder in inference mode to obtain predicted embeddings. 
Given a batch $\{(\mathbf{x}_i, \mathbf{y}_i, k_i)\}_{i=1}^N$, where $k_i$ is the subject index, we compute $\hat{\mathbf{y}}_i = f_\theta(\mathbf{x}_i; k_i),$
and collect all predictions $\{\hat{\mathbf{y}}_i\}$ and corresponding ground-truth targets $\{\mathbf{y}_i\}$ for retrieval-based evaluation.

Evaluation is performed within subject to factor out inter-subject variability. For each subject $s$, we have a \emph{query set} $\mathcal{Q}_s=\{\hat{\mathbf{y}}^{(s)}_i\}_{i=1}^{n_s}$ (predicted embeddings) and a \emph{reference set} 
$\mathcal{R}_s=\{\mathbf{y}^{(s)}_j\}_{j=1}^{n_s}$ (ground-truth embeddings). 
Each query $\hat{\mathbf{y}}^{(s)}_i$ has a unique paired target $\mathbf{y}^{(s)}_i$ in the reference set.
Correlation is measured with the cosine similarity, and the corresponding cosine distance (to be minimized) is $d_{\cos}(\hat{\mathbf{y}},\mathbf{y}) \;=\; 1 - \operatorname{cos}(\hat{\mathbf{y}},\mathbf{y}).$

For each subject $s$, we perform a nearest-neighbor search within $\mathcal{R}_s$ using cosine distance. 
Concretely, for each query $\hat{\mathbf{y}}^{(s)}_i$ we compute all pairwise distances
$d^{(s)}_{ij}\;=\;d_{\cos}\!\left(\hat{\mathbf{y}}^{(s)}_i,\;\mathbf{y}^{(s)}_j\right), \qquad j=1,\dots,n_s,$
rank reference embeddings by ascending distance, and select the $k$ closest matches:
$
\Pi^{(s)}_i(k) \;=\; \big\{\, j_1,\ldots,j_k \,\big\}\quad\text{with}\quad
d^{(s)}_{i j_1} \le \cdots \le d^{(s)}_{i j_k}.
$

Let $j^\star=i$ denote the index of the correct target for query $i$. 
The Top-$k$ accuracy for subject $s$ is defined as
\[
\text{Top-}k(s) \;=\; \frac{1}{n_s}\sum_{i=1}^{n_s} 
1[\, j^\star \in \Pi^{(s)}_i(k) \,\big],
\]
i.e., the fraction of queries for which the true target appears among the $k$ nearest neighbors. 
Overall performance is reported as the average across all subjects:
\[
\text{Top-}k \;=\; 
\frac{\sum_{s} \sum_{i=1}^{n_s} 1[\, j^\star \in \Pi^{(s)}_i(k) \,\big]}
{\sum_{s} n_s}.
\]
We report Top-1 and Top-3 accuracies for each stimulus modality: \textcolor{black}{Intuitively, Top-1 measures whether the brain-predicted embedding is closest to the correct stimulus out of all test candidates. Top-3 relaxes this to whether the correct stimulus appears among the three nearest neighbors, providing a measure of semantic proximity even when exact retrieval fails.} Given the sample size of the test set the chance level is Top-1: 1/250=0.40\% and Top-3: 3/250=1.20\% for HUTH dataset, Top-1: 1/980=0.10\% and Top-3: 3/980=0.31\% for NSD, Top-1: 1/60=1.67\% and Top-3: 3/60=5.00\% for GTZAN.

\section{Results}

\begin{table}[h]
\centering
\caption{Retrieval accuracies (mean $\pm$ std) per dataset/metric.}
\vspace{0.5em} 
\label{tab:panel_results}
\begin{tabular}{lllc}
\toprule
\textbf{Dataset} & \textbf{Metric} & \textbf{Method} & \textbf{Accuracy} (\%) \\
\midrule
\multirow{3}{*}{NSD (Image)} 
& \multirow{3}{*}{Top-1} 
    & Ridge Reg.      & $15.79 \pm 0.89$ \\
&   & Linear CL       & \textbf{$24.31 \pm 1.09$} \\
&   & Non-Lin. CL   & $20.87 \pm 2.12$ \\
\cmidrule(lr){2-4}
& \multirow{3}{*}{Top-3} 
    & Ridge Reg.      & $29.53 \pm 1.57$ \\
&   & Linear CL       & \textbf{$43.58 \pm 1.42$} \\
&   & Non-Lin. CL   & $39.51 \pm 1.93$ \\
\midrule
\multirow{3}{*}{HUTH (Lang.)} 
& \multirow{3}{*}{Top-1} 
    & Ridge Reg.      & $29.11 \pm 3.23$ \\
&   & Linear CL       & \textbf{$42.04 \pm 2.19$} \\
&   & Non-Lin. CL   & $38.23 \pm 2.68$ \\
\cmidrule(lr){2-4}
& \multirow{3}{*}{Top-3} 
    & Ridge Reg.      & $51.33 \pm 2.24$ \\
&   & Linear CL       & \textbf{$66.25 \pm 2.87$} \\
&   & Non-Lin. CL   & $61.70 \pm 1.83$ \\
\midrule
\multirow{3}{*}{GTZAN (Music)} 
& \multirow{3}{*}{Top-1} 
    & Ridge Reg.      & $22.67 \pm 1.56$ \\
&   & Linear CL       & \textbf{$33.13 \pm 1.47$} \\
&   & Non-Lin. CL   & $25.39 \pm 1.11$ \\
\cmidrule(lr){2-4}
& \multirow{3}{*}{Top-3} 
    & Ridge Reg.      & $49.10 \pm 2.00$ \\
&   & Linear CL       & \textbf{$57.97 \pm 1.12$} \\
&   & Non-Lin. CL   & $52.30 \pm 1.60$ \\
\bottomrule
\end{tabular}
\end{table}

\begin{figure*}
    \centering
    \includegraphics[width=.99\linewidth]{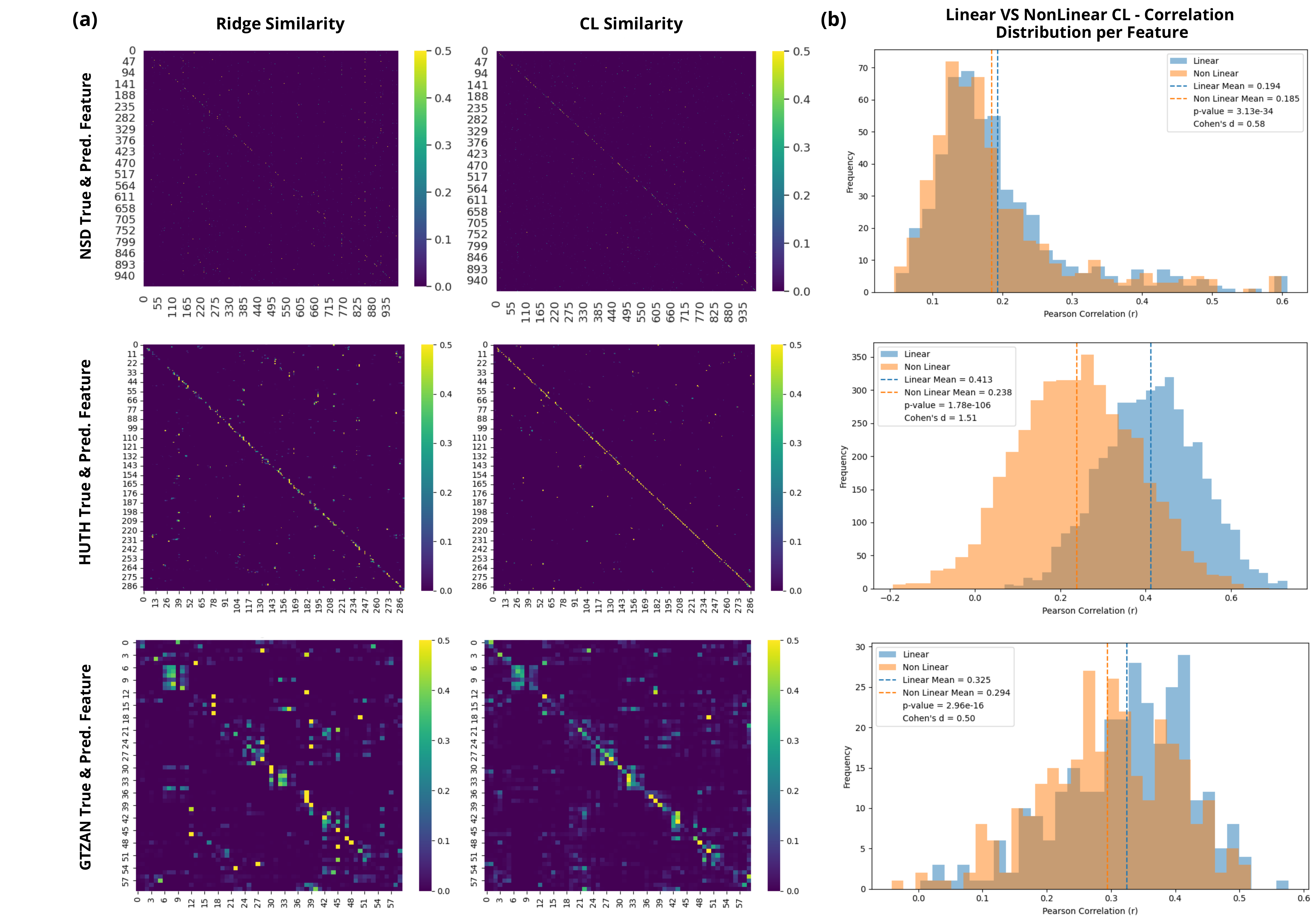}
    \caption{\textbf{(a)} Each heatmap represents the cosine similarity between predicted and ground-truth stimulus embeddings, computed sample-by-sample. Results are shown for three datasets (NSD, HUTH, GTZAN) and two models: a linear Ridge regression (left column) and the best contrastive learning model (the linear one, right column). The diagonal reflects correct predictions with high similarity between corresponding stimuli, while off-diagonal values indicate confusion between different candidates.
    All similarity matrices are normalized using a row-wise softmax to emphasize alignment between prediction and target embeddings. The CL model produces more concentrated diagonal patterns, indicating superior matching accuracy compared to Ridge approach.  
    \textbf{(b)} Each plot displays the distribution of Pearson correlation coefficients computed between model predictions (Linear in blue, Non-Linear in orange) and the ground truth stimulus embeddings, evaluated separately for each embedding feature. Dashed vertical lines indicate the mean correlation for each model. A t-test was performed for each comparison, testing the alternative hypothesis that the linear model yields higher correlations than the non-linear model. The resulting p-value and effect size (Cohen’s d) are reported in the legend. Results demonstrate that the linear model consistently achieves significantly higher correlations, with effect sizes ranging from moderate (Cohen’s d = 0.50) to large (Cohen’s d $>$ 1), depending on the dataset.}
    \label{fig:panel_results}
\end{figure*}

\begin{figure*}
    \centering
    \includegraphics[width=.90\linewidth]{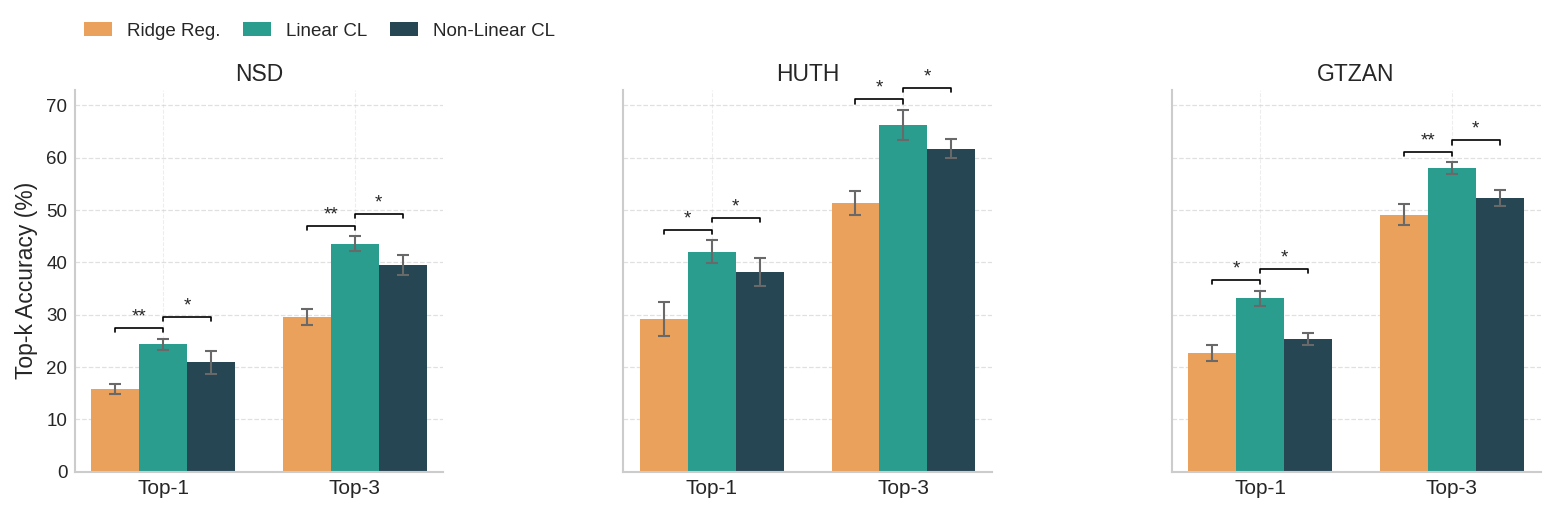}
    \caption{Quantitative bar charts to visualize decoding results. Stars above the bars reveal significance, according to the table in the Appendix. Double stars indicate pvalue lower than 1e-10.}
    \label{fig:visible_results}
\end{figure*}

Across all three modalities, decoding performance shows a consistent advantage for the linear contrastive model (Table~\ref{tab:panel_results} \& Figure \ref{fig:visible_results}). 
In the visual domain, linear contrastive learning achieved the highest retrieval accuracies, clearly outperforming both ridge regression and the non-linear variant. A similar trend was observed in the language domain, where the linear contrastive model provided the largest gains, with improvements particularly pronounced at the Top-3 level. In the musical domain, the same model again yielded superior accuracies, indicating that the benefits of contrastive learning extend beyond a single modality. This improvement is also qualitatively evident in Figure~\ref{fig:panel_results} (left panel), where correct embedding pairs (real and brain-predicted) are marked by sharper diagonal activations in the similarity matrices.

Comparisons between linear and non-linear mappings further demonstrate that architectural complexity does not translate into performance gains. 
Despite introducing additional parameters and activation functions (Table \ref{tab:hp_sweep_and_bests} in Appendix) between subject aligner and mapping layer, non-linear models consistently underperformed compared to the linear contrastive approach. Feature-wise evaluation (Figure~\ref{fig:panel_results}, right panel) confirmed that linear mappings lead to stronger correlations between predicted and ground-truth embeddings, suggesting that the critical factor is the contrastive objective itself rather than model complexity.

\begin{figure*}[h]
    \centering
    \includegraphics[width=1\linewidth]{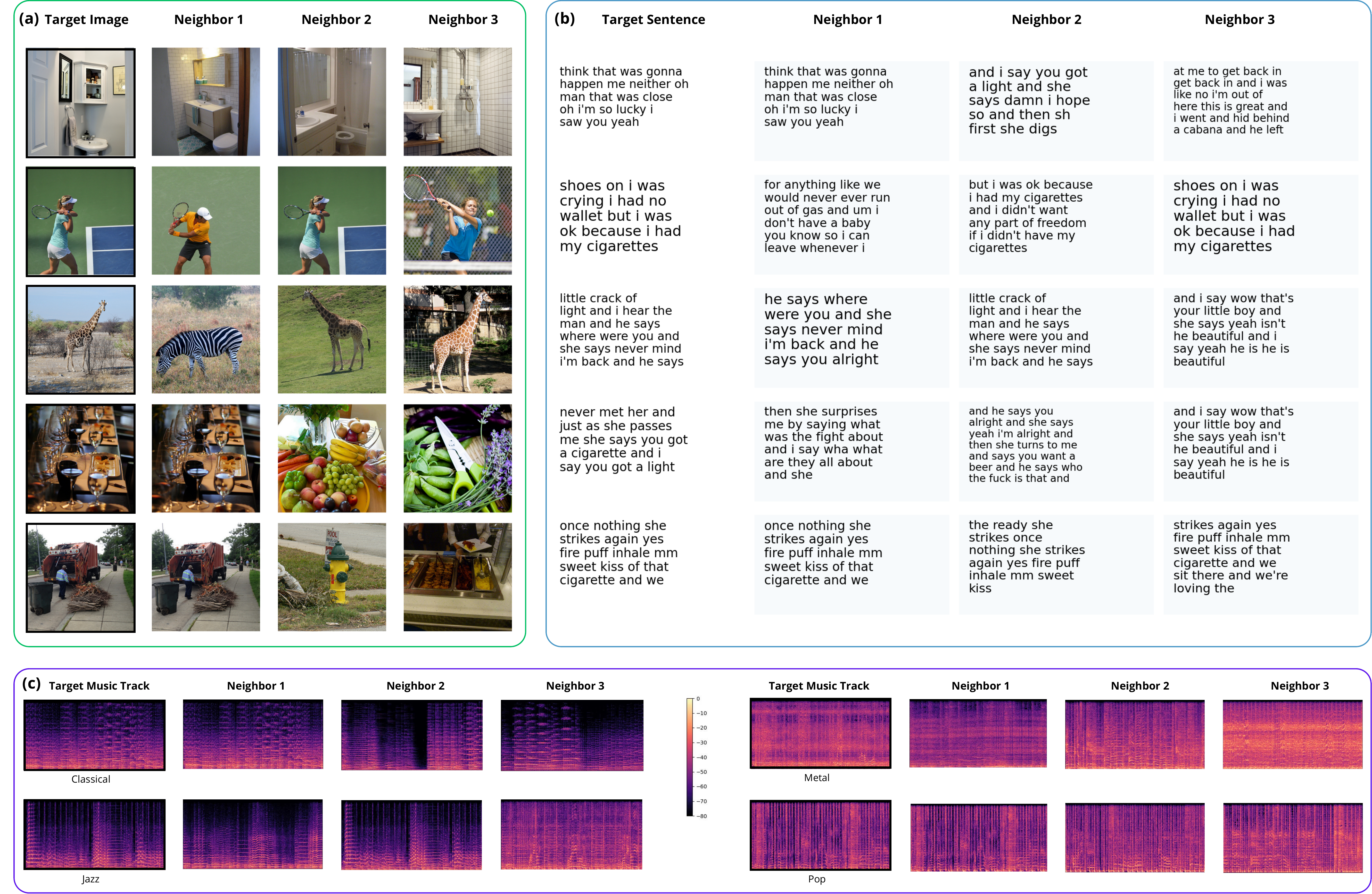}
    \caption{Random samples of brain decoding results. For each panel, the target column shows the ground-truth stimulus (music track, image, or sentence, depending on the modality), while the neighbor columns display the top retrieved candidates from the model’s latent space based on cosine similarity. (\textbf{a}) Retrieval of images viewed by participants. (\textbf{b}) Retrieval of text/sentences corresponding to the neural response. (\textbf{c}) Retrieval of music tracks from brain activity. These qualitative examples illustrate that the predicted neural embeddings often retrieve semantically related stimuli, highlighting the model’s ability to capture meaningful structure in brain representations.}
    \label{fig:panel_retrieved}
    \vspace{8mm}
\end{figure*}

In Figure \ref{fig:panel_retrieved}, we present qualitative decoding results from the test sets of the three datasets. In all cases, the retrieved samples show clear conceptual similarity with the target stimulus. For the NSD dataset, the retrieved images capture semantic content consistent with the reference, such as animals, food, or sports. A similar trend is observed in the textual modality, where the retrieved sentences convey the same high-level meaning as the ground-truth sentences. 
Finally, for the music dataset, the comparison of spectrograms highlights modality-specific correspondences: for instance, in the jazz genre case, the retrieved samples share distinctive frequency patterns visible in the target spectrogram; in contrast, for the metal sample, the retrieved spectrograms display higher energy across frequencies, reflecting the different acoustic structure of the genre.

\begin{table}
\centering
\caption{Ablation of the subject-alignment layer $A_k$. Anatomical-CL denotes a model trained separately per subject without the alignment layer; Linear-CL is the full multi-subject model with learned $A_k$.}
\vspace{0.5em}
\label{tab:alignment_ablation}
\begin{tabular}{llcc}
\toprule
Dataset & Metric & Linear-CL (\%) & Anat.-CL (\%) \\
\midrule
NSD & Top-1 & $24.31 \pm 1.09$ & $18.8 \pm 1.47$ \\
              & Top-3 & $43.58 \pm 1.42$ & $36.5 \pm 1.88$ \\
\midrule
HUTH & Top-1 & $42.04 \pm 2.19$ & $37.2 \pm 2.55$ \\
                & Top-3 & $66.25 \pm 2.87$ & $61.9 \pm 2.81$ \\
\midrule
GTZAN & Top-1 & $33.13 \pm 1.47$ & $27.4 \pm 1.69$ \\
              & Top-3 & $57.97 \pm 1.12$ & $50.5 \pm 1.92$ \\
\bottomrule
\end{tabular}
\end{table}

We also evaluated the contribution of the subject-specific alignment matrices $A_k$. Removing the alignment layer corresponds to training a separate anatomical-only model per subject. This ablation reduces performance across all datasets (Table \ref{tab:alignment_ablation}), confirming the benefit of learning a shared functional space.

\section{Discussion}

Our results indicate that a linear mapping trained with a contrastive objective is a robust and general strategy for fMRI decoding across modalities. Below we discuss two central findings --- the superiority of contrastive learning over ridge regression, and the consistent advantage of linear over shallow non-linear mappings --- and offer mechanistic explanations consistent with prior literature.

\subsection{Contrastive Learning for Vectorial Representation of Concepts}
If concepts in the brain are organized as vectors in high-dimensional spaces \citep{whyconcepts2024}, then meaning is carried primarily by their relative geometry: distances, angles, and directions capture similarity, typicality, and relational structure. This view suggests that learning and comparison are fundamentally geometric operations \citep{ferrante2025evidence}. Contrastive objectives \citep{chen2020simpleframeworkcontrastivelearning} directly operationalize this principle by maximizing angular similarity between matched pairs while enforcing separation from distractors, effectively aligning the training loss with the retrieval metric. In practice, the negative set acts as a data-driven regularizer: it suppresses directions that reflect nuisance variance in fMRI but are not discriminative in the target space, while amplifying those aligned with semantic information. Our findings that contrastive mappings systematically outperform ridge regression support this interpretation. Whereas ridge minimizes point-wise $\ell_2$ error—implicitly prioritizing magnitude alignment—contrastive learning preserves and sharpens the relational structure of the embedding manifold. This suggests that geometry-aware objectives are better matched to the retrieval metric used here than pointwise regression losses.

Notably, the same advantage holds when mapping into embedding spaces that are not trained contrastively, such as \textit{LLaMA3} language embeddings. This indicates that the benefit is not merely ‘contrastive-to-contrastive alignment’, but a more general advantage of geometry-aware objectives in aligning fMRI to vector spaces.


\subsection{Linear vs Non-Linear Mapping}
A second central finding is the consistent superiority of linear over shallow non-linear mappings. While this may appear counterintuitive, given the assumption that non-linear networks are needed to capture neural complexity, large-scale analyses of fMRI show that linear models perform better at the macroscale \citep{Nozari2024, schulz2020different}. This apparent linearity has a principled explanation: although neurons are individually non-linear, fMRI signals reflect averages over millions of units, filtered in time and further smoothed by observation noise. These operations suppress higher-order dynamics and yield an effective signal that approximates a first-order (Taylor-linear) expansion of the underlying neural processes. In this setting, linear mappings are not just a simplifying choice but may be the most explainable and faithful representation of the observable data.
Our experiments are consistent with this view: non-linear layers, while increasing expressivity, also relax the inductive bias that preserves embedding geometry. In the small-to-moderate data regime typical of fMRI, this flexibility can rotate or distort informative directions, amplify noise-driven variance, and disrupt calibration of vector norms, ultimately degrading retrieval accuracy. Of course, the space of possible non-linear architectures is vast, and we cannot exclude that specific designs or training regimes may close this gap.


Taken together, these perspectives highlight a broader principle: when decoding with rich, pre-trained representations, most of the relevant non-linear structure is already embedded upstream in the foundation models. The decoder’s role is not to discover new features but to align noisy brain measurements with an existing embedding geometry. 
Linear contrastive mappings are therefore well suited: they provide stable optimization, suppress nuisance variance, and maximize discriminative alignment. Conceptually, they instantiate two converging ideas: that semantic information in the brain is organized in vector spaces, and that fMRI provides a linearized view of these representations. 

\subsection{Subject-Alignment Layer}
Generally, fMRI responses exhibit strong inter-subject variability in both anatomical organization and functional topography. As shown in recent cross-subject decoding frameworks \citep{tang2025semantic,ferrante2023eyes,d2025tribe,aggarwal2024across,thual2023aligning}, anatomical alignment alone is insufficient for high-level semantic decoding; a subject-specific functional alignment is typically required to map different brains into a shared representational space. Our alignment layer $A_k$ plays exactly this role. Rather than being applied as a separate preprocessing step, it is jointly optimized inside the contrastive objective, allowing the model to learn linear subject-specific transformations (matrix) that project each subject’s fMRI activity into a common functional space in which stimulus representations are comparable.

\subsection{Limitations}
Several limitations of this work should be acknowledged. First, while our analyses systematically compared linear, ridge, and shallow non-linear decoders, the space of non-linear architectures is large. It remains possible that alternative designs or more extensive hyperparameter sweeps could get better performance. However, this consideration also reinforces our main claim: in practice, the computational cost of exhaustively searching for an optimal non-linear configuration may not be justified, since strong performance can already be achieved with simple linear contrastive approaches. Additionally, we did not evaluate time-aware architectures (e.g., LSTMs, Transformers) since temporal structure has been removed by design (GLM betas and HRF-averaged). Second, our study does not directly address generalization across datasets. Decoding models were trained and evaluated within individual modalities, leaving open the question of how well such mappings transfer across datasets. 
\textcolor{black}{Third, any possible "heavy lifting" done by the foundation model is constant across conditions and cannot explain the relative advantage of the linear contrastive decoder: the non-linear structure relevant to the task is already encoded upstream in the foundation model, and the decoder's role is purely one of geometric alignment.}
Finally, decoding brain activity into rich semantic spaces raises concerns about potential misuse, especially if applied to unconstrained settings or without informed consent. Future work should be guided not only by scientific objectives but also by principled discussions of data governance, individual rights, and ethical safeguards \citep{yuste2017four}.


\section{Conclusions}
\label{sec:conclusions}

We presented a unified framework for fMRI decoding that maps neural responses into the embedding spaces of large foundation models and we evaluated it across three distinct modalities (vision, language, and music) using the same pipeline. Empirically, a very simple strategy, a subject-aligned linear mapping trained with a contrastive objective (NT-Xent), consistently outperforms both ridge regression and non-linear MLP decoders. The core contribution of this work therefore lies less in algorithmic novelty and more in the strength of the evidence and the methodological message it supports: \emph{``do not overcomplicate, contrastive alignment with a linear decoder often suffices''}. We do not claim state-of-the-art performance on retrieval: our experiments compare decoder depth and objectives under a common framework. While our study focuses on retrieval-based decoding, the results provide a strong, motivated and reproducible approach for future work that may extend contrastive-linear alignment to generative reconstruction, cross-modality transfer, and integration with higher-resolution neural measurements. 

Our central claim is not that linear models are universally optimal, but that under standard fMRI preprocessing and matched data budgets, the choice of loss function (contrastive vs MSE) has a far greater impact on decoding performance than architectural depth.


\bibliographystyle{unsrt}
\bibliography{main}

\appendix
\setcounter{table}{0}
\setcounter{figure}{0}
\renewcommand{\thetable}{A\arabic{table}}
\renewcommand{\thefigure}{A\arabic{figure}}
\clearpage

\section{Statements}

\section*{Ethics Statement}
This study uses only publicly available datasets. No new human-subject data were collected. The original datasets were acquired with informed consent and institutional ethical approval. 

\section*{Data \& Code Availability}
All datasets used in this study are publicly available. The Natural Scenes Dataset (NSD) is available at \\
\url{https://naturalscenesdataset.org/} and through the AWS Open Data Registry at \url{https://registry.opendata.aws/nsd/}; access requires completion of the NSD Data Access Agreement. The natural language fMRI dataset used for the HUTH experiments is available on OpenNeuro at \url{https://openneuro.org/datasets/ds003020/versions/3.1.1}. The Music Genre fMRI Dataset used for the GTZAN experiments is available on OpenNeuro at \url{https://openneuro.org/datasets/ds003720/versions/1.0.1}.

All experiments can be reproduced with the scripts provided as github repository at \url{https://anonymous.4open.science/r/fMRIdatasets-decoding-3D12/}.

\section*{Author Contributions}
Matteo Ciferri: Conceptualization, Data curation, Formal analysis, Investigation, Methodology, Software, Visualization, Writing – original draft. Matteo Ferrante: Conceptualization, Formal analysis, Investigation, Methodology, Visualization. Nicola Toschi: Funding acquisition, Project administration, Resources, Supervision, Writing – review \& editing.

\section*{Competing interests}
The authors declare no competing interests.

\section*{Acknowledgements}
This work was supported by NEXTGENERATIONEU (NGEU) and funded by the Italian Ministry of University and Research (MUR), National Recovery and Resilience Plan (NRRP), project MNESYS (PE0000006) (to NT)– A Multiscale integrated approach to the study of the nervous system in health and disease (DN. 1553 11.10.2022); by the MUR-PNRR M4C2I1.3 PE6 project PE00000019 Heal Italia (to NT); by the NATIONAL CENTRE FOR HPC, BIG DATA AND QUANTUM COMPUTING, within the spoke "Multiscale Modeling and Engineering Applications" (to NT); the EXPERIENCE project (European Union’s Horizon 2020 Research and Innovation Programme under grant agreement No. 101017727); the CROSSBRAIN project (European Union’s European Innovation Council under grant agreement No. 101070908).

\section{Supplementary Material}

\subsection{Statistical significance and effect sizes}\label{secA1}

We quantified the statistical significance of the improvements reported in Table~1 of the main paper. For each dataset and metric, we performed paired $t$-tests across subjects and random seeds, comparing the Linear-CL decoder against both Ridge regression and the non-linear MLP. Effect sizes are reported as paired Cohen's $d$. See table \ref{tab:ttests}.

\begin{table}[h]
\centering
\caption{$t$-tests on Top-1 and Top-3 (L = Linear-CL, R = Ridge, N = Non-Linear).}
\label{tab:ttests}
\vspace{0.5em}
\setlength{\tabcolsep}{4pt}
\begin{tabular}{p{2.1cm}ccccc}
\toprule
Dataset & Metric & Comp. & $t$ & $p$ & $d$ \\
\midrule
NSD & Top-1 & L--R & 16.45 & $2.1{\times}10^{-12}$ & 2.69 \\
    & Top-3 & L--R & 18.91 & $5.7{\times}10^{-14}$ & 2.98 \\
    & Top-1 & L--N & 9.73 & $6.8{\times}10^{-7}$ & 1.46 \\
    & Top-3 & L--N & 11.14 & $3.2{\times}10^{-9}$ & 2.05 \\
\midrule
HUTH & Top-1 & L--R & 5.11 & $4.4{\times}10^{-4}$ & 1.23 \\
     & Top-3 & L--R & 6.78 & $2.6{\times}10^{-5}$ & 1.40 \\
     & Top-1 & L--N & 2.21 & $4.4{\times}10^{-2}$ & 0.59 \\
     & Top-3 & L--N & 3.70 & $2.3{\times}10^{-3}$ & 0.92 \\
\midrule
GTZAN & Top-1 & L--R & 7.74 & $5.2{\times}10^{-8}$ & 1.55 \\
      & Top-3 & L--R & 12.96 & $5.1{\times}10^{-13}$ & 2.79 \\
      & Top-1 & L--N & 8.36 & $1.4{\times}10^{-8}$ & 1.67 \\
      & Top-3 & L--N & 6.43 & $1.2{\times}10^{-6}$ & 1.28 \\
\bottomrule
\end{tabular}
\end{table}

\subsection{MSE across models}\label{secA2}

We also report (in Table \ref{tab:mse}) the Mean Squared Error (MSE) between predicted and ground-truth embeddings for all datasets. As expected, Ridge regression (explicitly optimized for MSE) achieves the lowest reconstruction error, while contrastive models exhibit higher MSE despite superior retrieval performance.

These results highlight that lower MSE does not necessarily translate into better retrieval, since the contrastive objective is invariant under global rescaling of embeddings and optimizes relative similarity rather than absolute reconstruction error.

\begin{table}
\centering
\caption{Mean Squared Error (MSE; mean $\pm$ std.\ across subjects) for Ridge, Non-Linear MLP, and Linear-CL decoders.}
\vspace{0.5em}
\label{tab:mse}
\begin{tabular}{llc}
\toprule
Dataset & Model & MSE (mean $\pm$ std.) \\
\midrule
HUTH (Language) & Ridge        & $1.02 \pm 0.02$ \\
                & Non-Linear   & $8.28 \pm 0.97$ \\
                & Linear-CL    & $33.17 \pm 1.62$ \\
\midrule
NSD (Image)     & Ridge        & $0.188 \pm 0.003$ \\
                & Non-Linear   & $0.216 \pm 0.002$ \\
                & Linear-CL    & $0.262 \pm 0.003$ \\
\midrule
GTZAN (Music)   & Ridge        & $0.00141 \pm 0.00013$ \\
                & Non-Linear   & $0.219 \pm 0.020$ \\
                & Linear-CL    & $0.779 \pm 0.090$ \\
\bottomrule
\end{tabular}
\end{table}

\subsection{Hyperparameter exploration}\label{secA3}

For completeness, we report the additional hyperparameter analyses (Table \ref{tab:hp_sweep_and_bests}). We use the standard NT-Xent sampling scheme, with one positive pair per anchor and all other items in the batch acting as negatives, as in SimCLR and CLIP. The temperature parameter $\tau$ was tuned separately for each dataset. The values reported correspond to the best temperature per dataset and were stable across random seeds. We also tested multiple batch sizes, observing that performance consistently improved for larger batches, consistent with the behavior of contrastive losses where a larger number of in-batch negatives improves the estimation of the objective and stabilizes optimization.

The best model with “Identity” as activation function is not intended to represent a separate non-linear architecture. It is in fact the same architecture used in our Linear-CL model: a stack of linear layers with Identity activations, which collapses to a linear transformation. This is precisely our model that achieves the best performance. Importantly, this "multi-layer linear network" corresponds to a low-rank factorization of the full linear weight matrix. In practice, this acts as an additional form of regularization: the transformation is still linear, but decomposed into smaller matrices. 

\begin{table*}[htbp]
\centering
\caption{Hyperparameter search space and best values per dataset. All datasets achieved the best performance with ``Identity'' activation function, effectively linearizing the model.}
\label{tab:hp_sweep_and_bests}

\vspace{0.5em}

\begin{tabular}{ll}
\toprule
\textbf{Hyperparameter} & \textbf{Values explored} \\
\midrule
Hidden dim.        & \{4096, 2048, 1024\} \\
Activation func.   & \{Identity, ReLU, GELU\} \\
Num. Layers        & \{1, 2, 5\} \\
Learning Rate      & \{1e-3, 1e-4, 1e-5\} \\
Dropout            & \{0.0, 0.3, 0.5\} \\
Temperature $\tau$ & \{0.02, 0.05, 0.10, 0.50\} \\
Weight Decay       & \{1e-3, 1e-4\} \\
Batch Size         & \{128, 512, 1024, 2048\} \\
\bottomrule
\end{tabular}

\vspace{1em}

\begin{tabular}{lccccccc}
\toprule
\textbf{Data} & \textbf{Drop.} & \textbf{Hidd.} & \textbf{NL} & \textbf{LR} & $\boldsymbol{\tau}$ & \textbf{WD} & \textbf{BS} \\
\midrule
NSD   & 0.3 & 1024 & 2 & 1e-4 & 0.02 & 1e-3 & 2048 \\
HUTH  & 0.0 & 2048 & 2 & 1e-3 & 0.05 & 1e-4 & 1024 \\
GTZAN & 0.0 & 1024 & 1 & 1e-3 & 0.10 & 1e-4 & 2048 \\
\bottomrule
\end{tabular}

\end{table*}

\begin{figure}
    \centering
    \includegraphics[width=.90\linewidth]{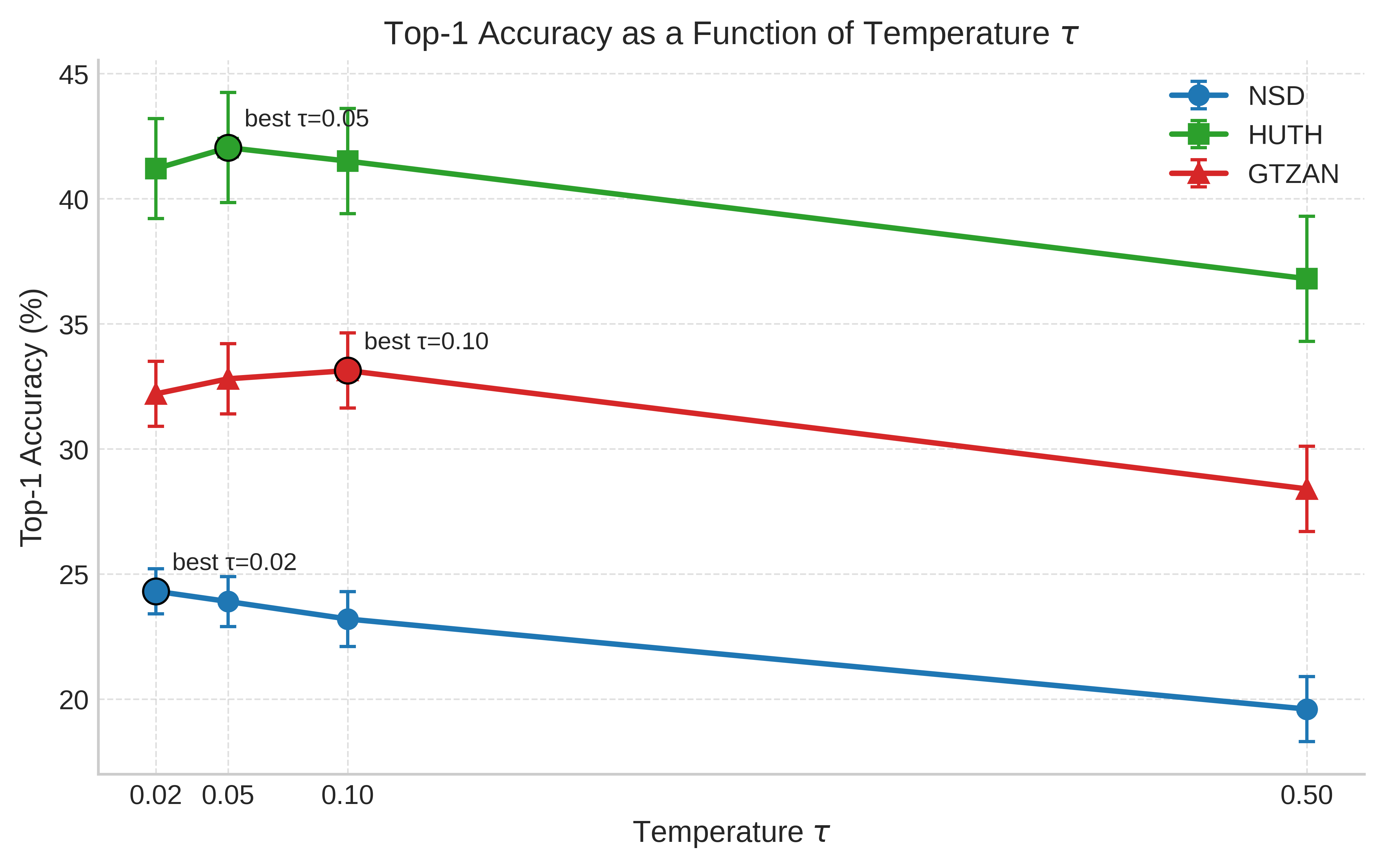}
    \caption{Top-1 accuracy as a function of contrastive learning temperature for all modalities.}
    \label{fig:tau_results}
\end{figure}

In Figure~\ref{fig:tau_results} we show Top-1 accuracy as a function of $\boldsymbol{\tau}$ in [0.02, 0.05, 0.10, 0.50] for all three datasets. Results show that performance is relatively stable in the range [0.02, 0.10] and degrades at $\boldsymbol{\tau}$ = 0.50, consistent with the behavior of NT-Xent where excessively high temperatures reduce the discriminative hardness of the contrastive task. The best $\boldsymbol{\tau}$ per dataset is already reported in Table \ref{tab:hp_sweep_and_bests}; the new figure provides the full sensitivity curve.

\subsection{Additional NSD Analyses}\label{secA4}

\paragraph{ROI-Level Decoding Performance.} 
To assess whether the advantage of the linear contrastive model is uniform across cortical regions, we evaluated Linear-CL and Non-Linear CL decoding accuracy separately across five functionally defined ROIs on the NSD dataset. Results are reported in Table~\ref{tab:roi}. The linear contrastive model consistently outperforms the non-linear variant across all regions. Notably, decoding performance is higher in higher-level semantic areas (PPA-OPA, EBA-FBA) than in early visual cortex (V1–V4-LO), consistent with encoding studies showing that CLIP embeddings are most strongly represented in semantic rather than structure-informative regions. \\

\begin{table}[h]
\centering
\caption{ROI-level retrieval accuracies on NSD for Linear-CL and Non-Linear CL decoders. Values are reported as mean $\pm$ std.}
\label{tab:roi}

\vspace{0.5em}
\setlength{\tabcolsep}{3pt}

\begin{tabular}{lcccc}
\toprule
\textbf{ROI} & \textbf{Lin. T1} & \textbf{Lin. T3} & \textbf{Non-Lin. T1} & \textbf{Non-Lin. T3} \\
\midrule
FFA-OFA   & $9.08 \pm 1.2$  & $16.99 \pm 1.8$ & $7.88 \pm 1.3$  & $15.87 \pm 1.9$ \\
PPA-OPA   & $13.81 \pm 1.4$ & $27.38 \pm 2.1$ & $12.79 \pm 1.5$ & $23.31 \pm 2.0$ \\
EBA-FBA   & $13.66 \pm 1.3$ & $26.13 \pm 2.0$ & $12.23 \pm 1.4$ & $24.25 \pm 1.9$ \\
VWFA      & $9.26 \pm 1.2$  & $18.70 \pm 1.7$ & $6.82 \pm 1.1$  & $16.84 \pm 1.6$ \\
V1--V4-LO & $6.38 \pm 1.1$  & $14.67 \pm 1.5$ & $5.83 \pm 1.0$  & $12.40 \pm 1.4$ \\
\bottomrule
\end{tabular}

\end{table}

\paragraph{SOTA comparison.} 
In order to improve comparability with prior work, we additionally evaluated our models following the retrieval protocol described in Mind-Eye2~\citep{scotti2024mindeye2}, noting that their model includes additional generative components outside the scope of this work. For each test sample, cosine similarity is computed between the brain-predicted embedding and 299 randomly selected test embeddings (300 total including the correct counterpart), repeated 30 times to account for sampling variability. Under this protocol, our Linear-CL model achieves a Top-1 accuracy of $40.55\% \pm 5.80\%$, and $36.79\% \pm 5.51\%$ for the Non-Linear model, placing our lightweight linear approach competitive.

\end{document}